\input{amssymb.sty}
\documentstyle[aps,prl,floats,epsf,epsfig]{revtex}

\def\Tc{T_{\mbox{\scriptsize c}}}
\def\rhoc{\rho_{\mbox{\scriptsize c}}}
\begin{document}

\twocolumn[\hsize\textwidth\columnwidth\hsize\csname@twocolumnfalse\endcsname 

\title{Precise Simulation of Near-critical Fluid Coexistence}

\author{Young C.\ Kim,$^{1}$ Michael E.\ Fisher,$^{1}$ and Erik Luijten$^{2}$}
\address{$^{1}$Institute for Physical Science and Technology, University of Maryland, College Park, Maryland 20742\\
$^{2}$Department of Materials Science and Engineering, University of Illinois at Urbana-Champaign, Urbana, Illinois 61801}

\date{\today}

\maketitle
\begin{abstract}
We present a novel method to derive liquid-gas coexisting densities, $\rho^{\pm}(T)$, from grand canonical simulations (without knowledge of $\Tc$ or criticality class). The minima of $ Q_{L}\equiv\langle m^{2} \rangle_{L}^{2}/\langle m^{4}\rangle_{L}$ in an $L$$\times$$L\times$$L$ box with $m = \rho - \langle\rho\rangle_{L}$ are used to generate recursively an unbiased universal finite-size scaling function. Monte Carlo data for a hard-core square-well fluid and for the restricted primitive model electrolyte yield $\rho^{\pm}$ to $\pm 1$-$2\%$ of $\rhoc$ down to 1 part in $10^4$-$10^3$ of $\Tc$ (and confirm well Ising character). Pressure mixing in the scaling fields is unequivocally revealed and indicates Yang-Yang ratios $R_{\mu} = -0.04_{4}$ and $0.2_{6}$ for the two models, respectively.\\
\\
$~~~~~~~~~~~~~~~~~~~~~~~~~~~~~~~~~~~~~~~~~~~~~~~~~~~~~$ PACS~ numbers:~ 64.60.Fr,~ 02.70.Rr,~ 05.70.Jk,~ 64.70.Fx
\end{abstract}

\vspace{0.2in}

]

Determining phase boundaries, critical points, and universality classes for various models that lack a clear symmetry has presented a serious difficulty in computer simulations \cite{ref1,ref2}. To tackle this problem, understanding scaling behavior in systems of finite size is crucial. However, as recently stressed \cite{ref3}, an important issue arises for {\em asymmetric} fluid criticality, even in the thermodynamic limit, namely, the potential presence of a Yang-Yang anomaly, in which the second derivative of the chemical potential, $\mu_{\sigma}(T)$, on the gas-liquid phase boundary diverges when the critical point, $\Tc$, is approached from below. To describe a Yang-Yang anomaly requires {\em pressure mixing} in the scaling fields \cite{ref3,ref4,ref5}. This also generates a term varying as $|t|^{2\beta}$ [with $t\,$$\equiv\,$$(T-\Tc)/\Tc$] in the gas-liquid coexistence diameter, that dominates the previously recognized $|t|^{1-\alpha}$ term \cite{ref6} and further distorts coexistence curves near criticality.

Our aim here is to show how coexistence curves may be estimated precisely and reliably near asymmetric critical points using grand canonical simulations, and to check our current understanding of scaling in such cases \cite{ref4,ref5}. It transpires that a finite-size scaling analysis at $\Tc$ also elucidates pressure mixing and allows us to measure its strength using simulation data.

\begin{figure}[h]
\vspace{-0.95in}
\centerline{\epsfig{figure=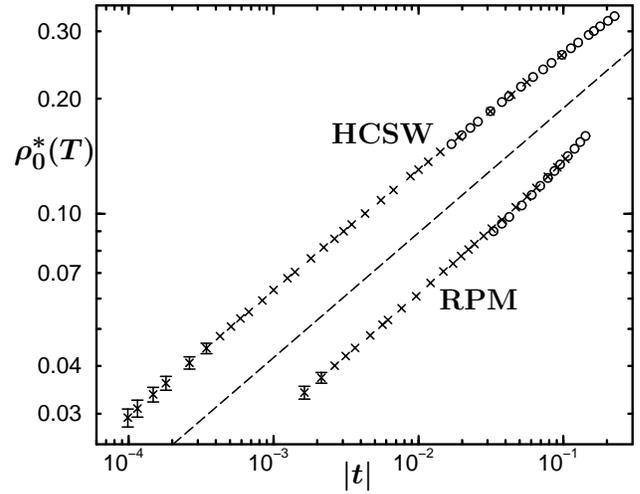,width=3.8in,angle=0}}
\vspace{-1.1in}
\caption{A log-log plot of the reduced semi-density-jump, $\rho_{0}^{\ast}\,$$=\,$$\frac{1}{2}[\rho^{+}(T)-\rho^{-}(T)]a^{3}$ vs.\ $t\,$$\equiv\,$$(T-\Tc)/\Tc$, where $a$ is the hard-sphere diameter, for a HCSW fluid with interaction range $1.5a$ (and $\rhoc^{\ast}\,$$\simeq\,$$0.3067$) [7] and for the RPM with $\rhoc^{\ast}\,$$\simeq\,$$0.079$ (at a $\zeta\,$$=\,$$5$ fine-discretization level [8]). The circles report previous estimates for the RPM and HCSW fluid [7] employing an equal-weight prescription [10]. The dashed line has a slope $\beta_{\mbox{\scriptsize Ising}}\,$$=\,$$0.32_{6}$. \label{fig1}}
\end{figure}

Figure \ref{fig1} presents our estimates of $\Delta\rho_{\infty}(T)\,$$\equiv\,$$\rho^{+}$$\,-\,$$\rho^{-}$,
 the density discontinuity across the phase boundary, for a hard-core square-well (HCSW) fluid and for the restricted primitive model (RPM) electrolyte, where $\rho^{+}(T)$ and $\rho^{-}(T)$ are the coexisting densities of liquid and vapor. The crosses represent new estimates obtained as explained below, while the open circles were derived previously directly from the observed double-peaked structure of the density distribution function in a finite grand canonical ensemble \cite{ref7}. Evidently the new approach yields estimates of $\rho^{+}(T)$ and $\rho^{-}(T)$ of precision $\pm1$-$2\%$ of $\rhoc$ or better, for temperatures $1.5$ to $2.5$ decades closer to the critical point. These results confirm convincingly that both models belong (as now expected \cite{ref7,ref9}) to the same $(d$=$3)$-dimensional Ising universality class: see below and the dashed line in Fig.\ \ref{fig1}.

To outline the established situation \cite{ref10}, recall that for $T\,$$<\,$$\Tc$ the grand canonical equilibrium distribution of the density, ${\cal P}_{L}(T;\rho)$, in a finite system of dimensions $L^{d}$ with periodic boundary conditions, has two Gaussian peaks near $\rho^{\pm}(T)$ when $L\,$$\gg\,$$a$, where $a$ measures the particle size. For large $L$ the two peaks are clearly separated and thus provide reasonable estimates for the coexisting densities via the equal-weight prescription \cite{ref11} --- the open circles in Fig.\ \ref{fig1} \cite{ref7}. However, when $\Tc$ is approached, finite-size effects, arising from the divergence of the correlation length, soon blur the distinction between the vapor and liquid states thereby seriously hampering the reliable estimation of the coexistence curve. An alternative procedure applicable near $\Tc$ is thus imperative.

Accordingly, we study the finite-system parameter $Q_{L}$ defined \cite{ref10,ref12,ref9} by the dimensionless moment-ratio
 \begin{equation}
   Q_{L}(T;\langle\rho\rangle_{L}) \equiv \langle m^{2}\rangle^{2}_{L}/\langle m^{4}\rangle_{L}, \hspace{0.1in} m = \rho-\langle\rho\rangle_{L},
 \end{equation}
where $\langle\cdot\rangle_{L}$ denotes a grand canonical expectation value at fixed $T$ and $\mu$. As well known, $Q_{L}\,$$\rightarrow\,$$\frac{1}{3}$ when $L\,$$\rightarrow\,$$\infty$ in any single-phase region of the $(\rho,T)$ plane while $Q_{L}\,$$\rightarrow\,$$1$ on the coexistence diameter, $\rho_{\mbox{\scriptsize diam}}(T)\,$$\equiv\,$$\frac{1}{2}(\rho^{+} + \rho^{-})$. At criticality, $Q_{L}$ rapidly approaches a universal value $Q_{\mbox{\scriptsize c}}$ \cite{ref9,ref10,ref12}, e.g., $Q_{\mbox{\scriptsize c}}\,$$=\,$$0.623_{6}$ for $d$=$3$ Ising systems. The $Q$-loci, $\rho_{Q}(T;L)$, on which $Q_{L}$ attains isothermal maxima, have recently provided a route to estimating $\Tc$ and $\rhoc$ with unprecedented precision \cite{ref9,ref13}.

In the two-phase region it has been known, but little appreciated, for some time [10(a),13], that $Q_{L}(T;\rho)$ displays a surprising singular behavior when $L\,$$\rightarrow\,$$\infty$ \cite{ref15}. This is illustrated by the dashed-line plots in Fig.\ 2,
\begin{figure}[ht]
\vspace{-1.0in}
\centerline{\epsfig{figure=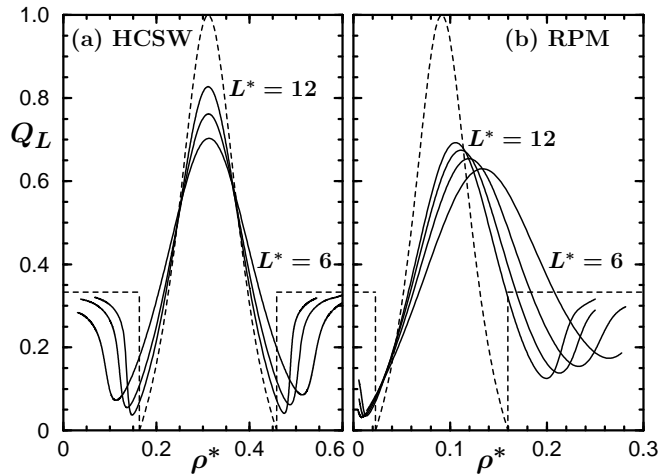,width=3.6in,angle=0}}
\vspace{-1.1in}
\caption{Plots of $Q_{L}(T;\langle\rho\rangle_{L})$ vs.\ $\rho^{\ast}\,$$\equiv\,$$\langle\rho\rangle a^{3}$ for (a) the HCSW fluid at $T^{\ast}\,$$=\,$$1.200$ $(<\,$$\Tc^{\ast}\,$$=\,$$1.2182_{1}$ [12]) and (b) the RPM at $T^{\ast}\,$$=\,$$0.0500$ $(<\,$$\Tc^{\ast}\,$$=\,$$0.0506_{9}$ [8]). The solid lines are for (a) $L^{\ast}\,$$\equiv\,$$L/a\,$$=\,$$6, 9, 12$ and (b) $L^{\ast}\,$$=\,$$6,8,10,12$; the dashed lines represent the {\em exact} limiting forms for the estimated values of $\rho^{+}$ and $\rho^{-}$ [15]. \label{fig2}}
\end{figure}
 which follow directly from the double-peaked structure of ${\cal P}_{L}(T;\rho)$ below $\Tc$ \cite{ref13,ref14,ref16} (together with our estimates for $\rho^{+}$ and $\rho^{-}$). Specifically, $Q_{\infty}(T;\langle\rho\rangle)$ exhibits a discontinuous drop from $Q_{\infty}\,$$=\,$$\frac{1}{3}$ to $Q_{\infty}\,$$=\,$$0$ on the two-phase boundaries, $\rho^{-}$ and $\rho^{+}$, and a continuous (but nonconvex \cite{ref13,ref16}) form for $\rho^{-}$$<\,$$\langle\rho\rangle\,$$<\,$$\rho^{+}$. For finite systems, however, the singularities are rounded and, as seen from the histogram-reweighted Monte Carlo simulations presented in Fig.\ 2, $Q_{L}(T;\langle\rho\rangle_{L})$ displays two smooth isothermal minima close to $\rho^{+}(T)$ and $\rho^{-}(T)$. It is notable that while the HCSW data are fairly symmetrical about $\rho_{\mbox{\scriptsize diam}}$, the RPM displays a remarkably strong asymmetry.

Clearly, it is tempting to extrapolate these minima in order to estimate $\rho^{+}(T)$ and $\rho^{-}(T)$ \cite{ref13}. However, when $\Tc$ is approached, naive extrapolation fails badly owing to the finite-size effects: indeed the graph of $Q_{L}(T;\langle\rho\rangle_{L})$ still exhibits two distinct minima {\em at} and {\em above} $T=\Tc$. Hence, some more powerful approach is necessary.

The behavior of $Q_{L}(T;\langle\rho\rangle_{L})$ near criticality can be understood via a recently developed `complete' scaling theory that explicitly encompasses pressure mixing \cite{ref3,ref4,ref5}. Specifically, the full thermodynamics of a one-component fluid near the bulk critical point $(p_{\mbox{\scriptsize c}},\Tc,\mu_{\mbox{\scriptsize c}})$ can be described with three relevant scaling fields
  \begin{eqnarray}
   & & \hspace{0.65in} \tilde{p} = \check{p}-k_{0}t -l_{0}\check{\mu} + \cdots,  \nonumber \\
   \tilde{t} & = & t - l_{1}\check{\mu} - j_{1}\check{p} + \cdots, \hspace{0.1in}  \tilde{h} = \check{\mu} - k_{1}t - j_{2}\check{p} + \cdots,  
  \end{eqnarray}
where the dimensionless deviations of the pressure and chemical potential from criticality are $\check{p}\,$$\equiv\,$$(p-p_{\mbox{\scriptsize c}})/\rhoc k_{\mbox{\scriptsize B}}\Tc,\,$ and $\,\check{\mu}\,$$\equiv\,$$(\mu - \mu_{\mbox{\scriptsize c}})/k_{\mbox{\scriptsize B}}\Tc$: the coefficients $j_{1}$ and $j_{2}$ measure the degree of pressure mixing, the Yang-Yang ratio $(\approx\,$$-T\mu_{\sigma}^{\prime\prime}/C_{V})$ being fixed by $R_{\mu}\,$$=\,$$-j_{2}/(1-j_{2})$ \cite{ref3,ref4}. For a finite box of dimensions $L^{d}$ with periodic boundary conditions, the finite-size scaling hypothesis now asserts \cite{ref4,ref5,ref13,ref17}
  \begin{equation}
   \rhoc\tilde{p} \approx L^{-d}Y(x,z), \hspace{0.1in} x = D\tilde{t}L^{1/\nu}, \hspace{0.1in} z = U\tilde{h}/|\tilde{t}|^{\Delta},
  \end{equation}
where we have used the hyperscaling relation $d\nu\,$$=\,$$2-\alpha$ (valid for $d\,$$<\,$$4$) and, for simplicity, neglected corrections to scaling. Note that $D$ and $U$ are {\em nonuniversal} amplitudes (of dimensions $L^{-1/\nu}$ and $L^{0}$, respectively), while $Y(x,z)$ is a {\em universal} function that is even in $z$ and independent of microscopic details while depending on the geometry and the boundary conditions of the system.

It follows that the full scaling expression for $Q_{L}$ is
  \begin{equation}
  {\cal Q}_{Q}(x,z)[ 1\hspace{-0.02in} +\hspace{-0.02in} A_{j}L^{-\kappa}{\cal Q}_{j}(x,z)\hspace{-0.02in} +\hspace{-0.02in} A_{l}L^{-\lambda} {\cal Q}_{l}(x,z)\hspace{-0.02in} + \cdots ],
  \end{equation}
\cite{ref13} with exponents and nonuniversal amplitudes
  \begin{eqnarray}
   \kappa & = & \beta/\nu, \hspace{0.25in} A_{j} = j_{2}D^{\Delta}U /\rhoc, \nonumber \\
   \lambda & = & (\Delta -1)/\nu,   \hspace{0.1in} A_{l} = (l_{1}+j_{1})D^{1-\Delta}/(1-j_{2}),
  \end{eqnarray}
while the scaling functions ${\cal Q}_{Q}$, ${\cal Q}_j$, and ${\cal Q}_l$ depend only on derivatives of $Y(x,z)$ thereby being {\em universal}. The symmetry of $Y(x,z)$ implies that ${\cal Q}_{Q}$ is even in $z$ while ${\cal Q}_j$ and ${\cal Q}_l$ are odd. Notice that the pressure mixing coefficient $j_{2}$ provides the dominant asymmetric $L$-dependent correction (with Ising values $\kappa\,$$=\,$$0.51_{7}\,$$<\,$$\lambda\,$$=\,$$0.89_{6}$) which, indeed, describes the strong asymmetric behavior of $Q_{L}(T;\langle\rho\rangle_{L})$ for the RPM seen in Fig.\ 2(b).

Of course, the mean density $\langle\rho\rangle_{L}$ also has a scaling form which we choose to write as \cite{ref13}
 \begin{eqnarray}
   y(T;L) & \equiv & 2[\langle\rho\rangle_{L} - \rho_{\mbox{\scriptsize diam}}(T)]/\Delta\rho_{\infty}(T) \nonumber \\
  & = & {\cal Y}\left[ 1 + A_{j}L^{-\kappa}{\cal Y}_{j} + A_{l}L^{-\lambda} {\cal Y}_{l} + \cdots \right],
  \end{eqnarray}
where, again, the scaling functions ${\cal Y}(x,z)$, ${\cal Y}_{j}(x,z)$, and ${\cal Y}_{l}(x,z)$ derive from $Y(x,z)$ and are universal, while ${\cal Y}$ is odd in $z$, and ${\cal Y}_{j}$ and ${\cal Y}_{l}$ are even.

The crucial point here is that $\rho_{\mbox{\scriptsize diam}}\,$$\propto\,$$(\rho^{+}+\rho^{-})$ and $\Delta\rho_{\infty}\,$$\propto\,$$(\rho^{+}-\rho^{-})$ embody the desired coexistence values $\rho^{+}(T)$ and $\rho^{-}(T)$. Our strategy will be to determine values for $\rho_{\mbox{\scriptsize diam}}$ and $\Delta\rho_{\infty}$ so that the minima of $Q_{L}(T;\langle\rho\rangle_{L})$, say, $Q_{\mbox{\scriptsize m}}^{+}(T;L)$ and $Q_{\mbox{\scriptsize m}}^{-}(T;L)$, and their locations, $\rho_{\mbox{\scriptsize m}}^{+}(T;L)\,$$>\,$$\rho_{\mbox{\scriptsize m}}^{-}(T;L)$, satisfy appropriate scaling relations. We focus first on $\Delta\rho_{\infty}$ and, to minimize the effects of asymmetry (arising from the mixing coefficients $j_{2}$, $j_{1}$ and $l_{1}$), we examine the mean and difference
  \begin{equation}
   \bar{Q}_{\mbox{\scriptsize min}} \equiv \mbox{$\frac{1}{2}$}(Q_{\mbox{\scriptsize m}}^{+} + Q_{\mbox{\scriptsize m}}^{-}), \hspace{0.1in} \Delta y_{\mbox{\scriptsize min}} \equiv \mbox{$\frac{1}{2}$}(y_{\mbox{\scriptsize m}}^{+} - y_{\mbox{\scriptsize m}}^{-}).
  \end{equation}

Now, on evaluating (4) and (6) at $z_{\mbox{\scriptsize min}}^{\pm}$ (which asymptotically fixes $Q_{\mbox{\scriptsize m}}^{\pm}$) and formally eliminating $x\,$$\propto\,$$tL^{1/\nu}$ between the resulting expressions, we see that $\bar{Q}_{\mbox{\scriptsize min}}(T;L)$ and $\Delta y_{\mbox{\scriptsize min}}(T;L)$ should be related in a way that, to the orders displayed, is {\em independent} of $T$ and $L$ and (up to the neglected corrections to scaling) reflects only the universality class of the critical system under consideration. {\em A priori} this class is unknown --- and, indeed, is to be determined. However, for any scalar order parameter the two-peaked, double-Gaussian structure of ${\cal P}_{L}(T;\rho)$ should be reproduced asymptotically when $L\,$$\rightarrow\,$$\infty$ at fixed $T\,$$<\,$$\Tc$. On this basis it is straightforward to calculate the universal relation for $\bar{Q}_{\mbox{\scriptsize min}}\,$$\rightarrow\,$$0$: we find [12(a)]
  \begin{equation}
   \Delta y_{\mbox{\scriptsize min}}(q) = 1 + \mbox{$\frac{1}{2}$}q + {\cal O}(q^{2}),\,\, q\equiv \bar{Q}_{\mbox{\scriptsize min}}\ln (4/e\bar{Q}_{\mbox{\scriptsize min}}),
  \end{equation}
which, to this order, is independent of any asymmetry.

\begin{figure}[ht]
\vspace{-1.0in}
\centerline{\epsfig{figure=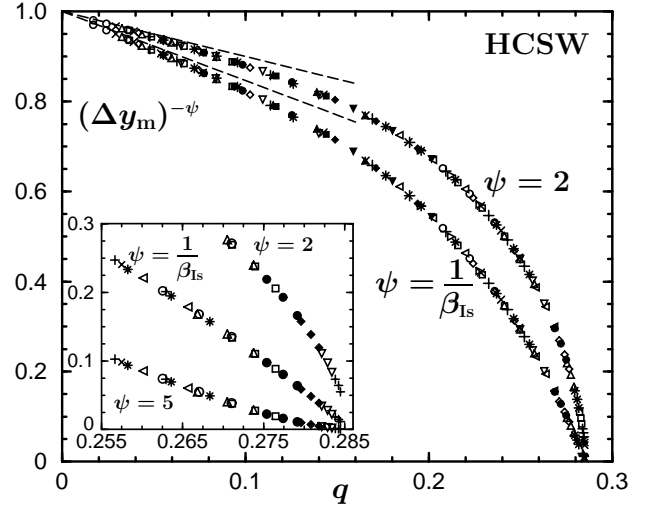,width=3.9in,angle=0}}
\vspace{-1.15in}
\caption{Scaling plots of $(\Delta y_{\mbox{\scriptsize min}})^{-\psi}$ (for $\psi$$\,=\,$$2$, $1/\beta_{\mbox{\scriptsize Is}}=3.0_{7}$, and $5$) vs.\ $q\,$$=\,$$\bar{Q}_{\mbox{\scriptsize min}}$$\ln (4/e\bar{Q}_{\mbox{\scriptsize min}})$ for the HCSW fluid built up recursively from low $q$ where the dashed lines are exact: see Eq.\ (8). Various symbols, most suppressed for clarity, depict results at increasing $T_{j}$: see text. \label{fig3}}
\end{figure}

Finally, we can employ our scaling analysis to generate the limiting coexistence curve recursively using finite-size simulation data for $Q_{L}$. Appropriate initial steps are: {\bf (i)} Collect data sets $\{ Q_{\mbox{\scriptsize m}}^{\pm}(T;L_{i}),\rho_{\mbox{\scriptsize m}}^{\pm}(T;L_{i})\}$ for a range of values $\{ L_{i} \}_{1}^{n}$ at fixed values of $T\,$$\lesssim\,$$\Tc$. {\bf (ii)} For a value $T$$\,=\,$$T_{0}$ sufficiently low that $\bar{Q}_{\mbox{\scriptsize min}}\,$$\lesssim\,$$0.03$ [which corresponds to well separated peaks in ${\cal P}_{L}(T_{0};\rho)$], choose a density-jump value, say $\Delta\rho_{T_{0}}$, independent of $i$, which leads to the best fit of $\Delta y_{\mbox{\scriptsize min}}^{(i)}\,$$\equiv\,$$[\rho_{\mbox{\scriptsize m}}^{+}(T_{0};L_{i})-\rho_{\mbox{\scriptsize m}}^{-}(T_{0};L_{i})]/\Delta\rho_{T_{0}}$ vs.\ $q_{0}^{(i)}\,$$\equiv\,$$q(T_{0};L_{i})$ to the relation (8) at small $q$: see the dashed lines in Fig.\ \ref{fig3}.
 In light of the scaling relations (4) and (6), the parameter $\Delta\rho_{T_{0}}$ can then be identified as an estimate for $\Delta\rho_{\infty}(T_{0})$. {\bf (iii)} Increase $T_{0}$ to $T_{1}$$\,=\,$$T_{0}+\Delta T_{0}$ by a small $\Delta T_{0}$, chosen so that the new set $\{ q_{1}^{(i)}\}^{n}_{1}$ overlaps the previous one. {\bf (iv)} Determine a new value, $\Delta\rho_{T_{1}}$, so that the plotted data display an optimal collapse that extends the previous numerical scaling function to larger values of $q$: see the gradual departure of the fits from the dashed lines in Fig.\ \ref{fig3} as $q$ increases. In practice we have found that $n\,$$=\,$$3$ distinct box sizes with $L_{3}\,$$\gtrsim\,$$1.3L_{1}$ may well suffice. {\bf (v)} Repeat steps {\bf (iii)} and {\bf (iv)} generating successive estimates for $\Delta\rho_{\infty}(T_{j})$ for $j$$\,=\,$$2$, $3$, $\cdots$. Smaller increments $\Delta T_{j}$ are needed as $T_{j}$$\rightarrow\,$$\Tc$ and the $q_{j}^{(i)}$ increase to $q_{\mbox{\scriptsize c}}\,$$=\,$$Q_{\mbox{\scriptsize min}}^{\mbox{\scriptsize c}} \ln (4/e Q_{\mbox{\scriptsize min}}^{\mbox{\scriptsize c}})$ (see Fig.\ \ref{fig3}) so that histogram-reweighting procedures are crucial \cite{ref7,ref9}.

Figure \ref{fig3} presents a scaling plot for the HCSW fluid constructed in this fashion: system sizes $L^{\ast}\,$$\equiv\,$$L/a\,$$=\,$$9,10.5,$ and $12$ were used and led to the estimates shown in Fig.\ 1 for $\Delta\rho_{\infty}(T)$ from $|t|\,$$\simeq\,$$0.23$ down to $|t|\,$$\simeq\,$$10^{-4}$. Purely for ease of presentation, Fig.\ \ref{fig3} displays $(\Delta y_{\mbox{\scriptsize min}})^{-\psi}$ for selected values of $\psi$. In fact, the scaling analysis indicates that $\Delta y_{\mbox{\scriptsize min}}(q)$ should diverge like $(q_{\mbox{\scriptsize c}} - q)^{-\beta}$ when $q\,$$\rightarrow\,$$q_{\mbox{\scriptsize c}}$ as $T$$\rightarrow\,$$\Tc$, with $q_{\mbox{\scriptsize c}}$ a universal value (depending on geometry and boundary conditions) [12(b)]. For the HCSW fluid with periodic boundary conditions we find $Q_{\mbox{\scriptsize min}}^{\mbox{\scriptsize c}}\,$$=\,$$0.110_{2}$. To lower precision, the RPM data yield the same scaling plots and value of $Q_{\mbox{\scriptsize min}}^{\mbox{\scriptsize c}}$ [12(b)]. On the other hand, the approximate scaling form proposed by Tsypin and Bl\"{o}te \cite{ref18} for ${\cal P}_{L}(\Tc;\rho)$ for ($d$=$3$) Ising models gives $Q_{\mbox{\scriptsize min}}^{\mbox{\scriptsize c}}\,$$\simeq\,$$0.117$, only $6\%$ higher than we observe. For ($d$=$2$) Ising models we estimate $Q_{\mbox{\scriptsize min}}^{\mbox{\scriptsize c}}\,$$\simeq\,$$0.28$ using data in \cite{ref11}.

Evidently, the choice of $\psi\,$$=\,$$1/\beta$ should yield a plot that intersects the $q$ axis linearly; indeed, for the Ising value, $\beta_{\mbox{\scriptsize Is}}\,$$=\,$$0.32_{6}$, this is so. But, we emphasize that this observation plays no role in the calculation of Fig.\ \ref{fig1}.

Clearly, uncertainties in choosing $\Delta\rho_{T_{j}}$, $\Delta\rho_{T_{j+1}}$, $\cdots$ in steps {\bf (ii)} and {\bf (iv)} will propagate. Well below $\Tc$ (where care must be taken to ensure two-phase {\em equilibrium}) we can fit the limiting behavior (8) with a precision of $\pm 1.0\%$ or better in $\Delta\rho_{T}/\rhoc$. The overall uncertainties then grow by factors of $5$-$10$ as $|t|$ decreases to $10^{-4}$ for the HCSW fluid and $10^{-3}$ for the RPM: see Fig.\ \ref{fig1}.

It is also remarkable that the $\Delta\rho_{\infty}(T)$ estimates imply values for $\Tc$. For the HCSW fluid we thus find $\Tc^{\ast}\,$$\simeq\,$$1.21821(2)$ which lies close to the upper confidence limit of the previous estimate $\Tc^{\ast}\,$$\simeq\,$$1.2179(3)$ [7]: see also [13(a)] Eq.\ (5.6). For the RPM we obtain $\Tc^{\ast}\,$$\simeq\,$$0.05069(2)$ which agrees precisely with Ref.\ \cite{ref9}. Explicit fits to $\Delta\rho_{\infty}(T)$, that allow for the leading correction terms, yield $\beta\,$$=\,$$ 0.324(10)$ for the HCSW fluid and $\beta\,$$=\,$$0.34(5)$ for the RPM, so providing independent, albeit weaker confirmation of the Ising behavior established using data confined to $T\gtrsim \Tc$ \cite{ref7,ref9}.

The scaling results (4) and (5) suggest that evidence for a pressure-mixing coefficient $j_{2}$ might be detected in finite-size data. Indeed, a detailed calculation [12(b)] of the asymmetry seen in the minima of $Q_{L}$ {\em at} $T\,$$=\,$$\Tc$ yields
  \begin{equation}
   \left({\cal A}_{\mbox{\scriptsize min}} \equiv \frac{Q_{\mbox{\scriptsize m}}^{+}-Q_{\mbox{\scriptsize m}}^{-}}{Q_{\mbox{\scriptsize m}}^{+} + Q_{\mbox{\scriptsize m}}^{-}}\right)_{\mbox{\scriptsize c}} = A_{j}c_{j}L^{-\kappa} + A_{l}c_{l}L^{-\lambda} + \cdots,
  \end{equation}
where $c_{j}$ and $c_{l}$ are universal numbers determined by expansion coefficients of $Y(0,z)$ about the minima at $z_{\mbox{\scriptsize min}}^{\pm}$. Recall from (5) that $A_{j}$ is proportional to $j_{2}$.

In Fig.\ \ref{fig4} we present data for ${\cal A}_{\mbox{\scriptsize min}}^{\mbox{\scriptsize c}}(L)$ for the RPM and the HCSW fluid:
\begin{figure}[ht]
\vspace{-1.05in}
\centerline{\epsfig{figure=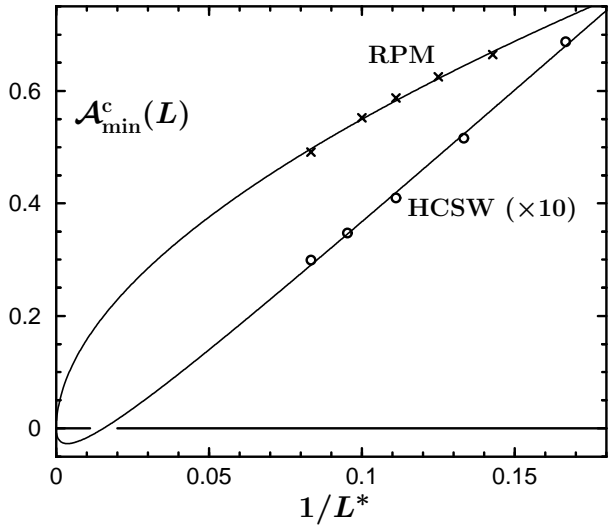,width=3.9in,angle=0}}
\vspace{-1.1in}
\caption{Plots of the critical asymmetry factor ${\cal A}_{\mbox{\scriptsize min}}^{\mbox{\scriptsize c}}(L)$: see Eq.\ (9). The fitted curves use Ising exponent values and indicate relatively large pressure mixing in the RPM. \label{fig4}}
\end{figure}
 even by eye, the former strongly suggest a leading exponent closer to $\kappa\,$$=\,$$0.51_{7}$ than to $\lambda\,$$=\,$$0.89_{6}$. The fits in Fig.\ \ref{fig4}, using only the two leading terms in (9), support this but also indicate a weak $j_{2}$ contribution of opposite sign for the HCSW fluid. Further fairly elaborate analysis \cite{ref13} yields $j_{2}\,$$=\,$$- 0.35(7)$, implying a strong, $R_{\mu}\,$$=\,$$0.26(4)$, Yang-Yang anomaly for the RPM, while $j_{2}\,$$=\,$$0.042(3)$ and $R_{\mu}\,$$=\,$$-0.044(3)$ for the HCSW fluid. The latter result is consistent with the earlier, much less precise estimate $R_{\mu}\,$$\simeq\,$$-0.08(12)$ [7].

Finally, to determine the diameter $\rho_{\mbox{\scriptsize diam}}(T)$ we compare $\bar{y}_{\mbox{\scriptsize min}}\,$$\equiv\,$$\frac{1}{2}(y_{\mbox{\scriptsize m}}^{+}+y_{\mbox{\scriptsize m}}^{-})$ and ${\cal A}_{\mbox{\scriptsize min}}(T;L_{i})$. Analysis of the two-Gaussian limit [12(b)] yields $\bar{y}_{\mbox{\scriptsize min}}/{\cal A}_{\mbox{\scriptsize min}}\,$$=\,$$\frac{1}{2}\bar{q} + {\cal O}(\bar{q}^{2})$ with $\bar{q}$$\,\equiv\,$$q-\bar{Q}_{\mbox{\scriptsize min}}$ which is again universal in leading order. Owing to the asymmetric terms in (4) and (6) the analogous scaling plots are now more sensitive to nonuniversal details and exhibit small, $L$-dependent corrections when $q$ approaches $q_{\mbox{\scriptsize c}}$. Nevertheless, the approach succeeds and the critical densities, $\rhoc^{\ast}$, predicted from the diameters when $T$$\rightarrow\,$$\Tc$ are fully consistent with the previous, $T\,$$\gtrsim \,$$\Tc$ estimates \cite{ref7,ref9,ref13}. Details for both the RPM and the HCSW fluid will be presented elsewhere [12(b)]. 

In summary, we have shown how the finite-size scaling information hidden in precise simulation data can be systematically extracted via a novel $Q$-minima recursive algorithm to yield coexistence curves far closer to $\Tc$ and with a much higher precision than previously appeared possible. As a byproduct, pressure mixing has been quantitatively resolved.

We are grateful to Dr.\ G.\ Orkoulas for help in generating $Q_{L}$ data for the hard-core square-well fluid, and to him, Prof.\ K.\ Binder, and Prof.\ A.\ Z.\ Panagiotopoulos for their interest.  The support of the National Science Foundation (through Grant No.\ CHE 99-81772 to M.E.F.) and of the Department of Energy (for E.L. via the F. Seitz Materials Research Laboratory under contract No.\ DEFG02-91ER45439) is gratefully acknowledged.


\begin{thebibliography}{99}
\bibitem{ref16} Note that the comment on $Q_{\infty}(T;\langle\rho\rangle)$ in Ref.\ 9 is misleading; see also [13(a)] Eq.\ (23) and [12(a)].
\end{thebibliography}
\end{document}